\documentclass[twocolumn,preprintnumbers,amssymb,amsmath,9pt]{revtex4}[10pt]
\usepackage{graphicx}
\usepackage{dcolumn}
\usepackage{bm}
\usepackage{amssymb}
\usepackage{epsfig}
\newcommand{\be}{\begin{equation}}
\newcommand{\ee}{\end{equation}}
\newcommand\beq{\begin{eqnarray}}
\newcommand\eeq{\end{eqnarray}} 
\newcommand\eqn[1]{\label{eq:#1}}

\newcommand{\GeV}{{\rm ~GeV }}
\newcommand{\TeV}{{\rm ~TeV }}

\newcommand{\cm}{{\rm ~cm}}

\newcommand{\gsim}{ \mathop{}_{\textstyle \sim}^{\textstyle >} }
\newcommand{\lsim}{ \mathop{}_{\textstyle \sim}^{\textstyle <} }

\begin{document}


\title{Slightly Non-Minimal Dark Matter in PAMELA, ATIC and Fermi}

\author{Ann E. Nelson$^1$}
\email{anelson@phys.washington.edu}
\author{Christopher Spitzer$^1$}
\email{cspitzer@u.washington.edu}

\affiliation{$^1$Dept. of Physics, University of Washington,  Seattle, WA 98195-1560,USA}  
  
\begin{abstract}
We present a  simple model in which dark matter couples to the standard model through a light scalar intermediary that is itself unstable.  We find that this model has several notable features, and allows a natural explanation for a surplus of positrons and electrons, but no surplus of anti-protons, as has been suggested by early data from PAMELA and ATIC.  Moreover, this model yields a very small nucleon coupling, well below the direct detection limits.  In this paper we explore the effect of this model in both the early universe and in the galaxy.
\end{abstract}

\date{\today}
\maketitle

\section{Introduction}

There is very strong evidence that nonbaryonic dark matter comprises about 21\% of the present energy density of the universe.  Identification of the dark components poses one of the strongest challenges in particle physics and astrophysics,  and will almost certainly require an extension to the standard model.  
Indeed, there is a rich literature on dark matter candidates that arise more or less naturally from various extensions.

A number of recent experiments have
reported observations that, if confirmed in their final analysis, may be the first strong non-gravitational hints to the identity of dark matter.  The PAMELA collaboration has reported that their satellite-born experiment has observed an excess in the ratio of positrons to the total expected electron plus positron background.  The spectrum they reported departed from background at 10 GeV and rose monotonically to 100 GeV\cite{Adriani:2008zr}.
Recently the PAMELA experiment has provided an updated spectrum that incorporates additional data and a new statistical method \cite{Adriani:2010ib}. We will use the ``beta-fit" from that paper.
Their final data release is expected to have the positron spectrum measured to  about 300 GeV \cite{Boezio:2008mp,Adriani:2008zr}.  
The balloon-based ATIC experiment measured the electron spectrum in the range 10 GeV to 2 TeV.  
Their spectrum, given in \cite{Chang:2008zzr},
seems to show a significant excess from 300 GeV to 600 GeV, with a peak at roughly 500 GeV.  This data is consistent with published data from another balloon experiment, PPB-BETS, which measured electrons up to about 1 TeV~\cite{Torii:2008xu}, and with the earlier hints of an excess at the HEAT experiment~\cite{Barwick:1997ig}.
The Fermi Space Telescope has released a combined electron and positron spectrum with excellent statistics across the ATIC band, from $20\GeV$ to $1\TeV$~\cite{Abdo:2009zk}.  Their measurement does not agree with excess found by ATIC.  The spectrum does appear to have an excess of events with a peak near $400\GeV$, though the authors emphasize that the spectrum may be fit with a simple power law.

A number of groups have studied the consistency of these experiments with various theories of dark matter.  Some groups have considered the signals generated by specific models, including Minimal Dark Matter~\cite{Cirelli:2008jk}  the two-component $N_{{\rm DM}}$MSSM model~\cite{Huh:2008vj}, supersymmetry \cite{Barger:2008su,Bergstrom:2008gr} and a model with a nonthermal production mechanism \cite{Fairbairn:2008fb}.  Other groups have instead fit the data to general models as well as specific models~\cite{Cholis:2008hb,Cirelli:2008pk}.   Recently, Arkani-Hamed et. al. have proposed an interesting dark matter model that takes advantage of the Somerfield mechanism to achieve enhanced annihilation~\cite{ArkaniHamed:2008qn}, following earlier work in refs. \cite{Cirelli:2008pk}.  Their model, as well as the model of  Pospelov et al.~\cite{Pospelov:2007mp,Pospelov:2008jd},    share some characteristics with the one discussed in this paper due to the inclusion of a light boson to generate the lepton spectrum.
Earlier work by Cholis et al.~\cite{Cholis:2008vb} studied a scenario in which dark matter annihilates through a light intermediary. A study by March-Russel et al.~\cite{MarchRussell:2008yu} considers a supersymmetric dark matter model with a similar mechanism of communication with the standard model.  
Some early work on which galactic signals are generated by particular models appears in~\cite{Zeldovich:1980st,Fargion:1994me,Fargion:1999ss}.
Many of these models associate the dark matter with a new sector containing multiple exotic new particles.

In this paper we present a dark matter model that explains the experimental observations in a simple way and differs   significantly  from the scenarios mentioned above.  We assume that there is a new sector of particles that interact weakly with standard model particles, but may have stronger interactions among themselves.  In the new sector we take the lightest stable particle to have no direct standard model coupling.  The particle may still annihilate into   standard model  particles via lighter, unstable exotic intermediary particles.  If the intermediary is sufficiently light, an excess of galactic electrons and positrons is generated while baryonic production is kinematically disallowed.

The model has several other novel features.  Since the dark and standard model sectors decouple early on, the temperature of the dark matter is generically different from the photon temperature at the time the relic abundance is formed.  This modifies the allowed parameter space of relic masses and couplings.  We also note that the dark-matter nucleon cross-section is much smaller than in many scenarios, which makes direct detection a challenging prospect.

In the following section we present a specific model which we regard as a   minimal existence proof   of the feasability of reproducing the PAMELA positron excess from a dark sector containing a heavy stable particle and a light unstable intermediary, without any Somerfield enhancement. The mechanism described can be embedded in a number of extensions to the Standard Model, including hidden valleys \cite{Strassler:2006im} and unparticles \cite{Georgi:2007ek}.  Next, we examine the early-universe behavior of the model and derive   the dark matter density today in terms of model parameters.  We then calculate the present-day electron and positrons fluxes that result from galactic dark matter annihilation, and show that they are consistent with PAMELA and ATIC.  
This model provides a reasonable fit to the shape of the Fermi spectrum, but produces more electrons than observed without a significant modification to the boost factor.
Finally, we will conclude with a  few comments on  some distinct features of this scenario.

\section{Model Setup}

Consider the Standard Model extended by a number of new fields that we will label $X$, $Y$ and $Z_1, \ldots, Z_n$ where $n\geq 0$.  The fields named $Z_i$   are unrelated to the Standard Model $Z$ boson.  For simplicity we will take all new fields to be scalars, though the model is can be easily modified to include fermionic dark matter or intermediaries.  We'll take the Lagrangian to have the form
\beq
\mathcal{L} &\supset& M^2 X^2 + m^2 Y^2 + \sum_i m_i^2 Z_i^2 \nonumber \\
 & & + \lambda X^2 Y^2 + \alpha Y e \bar{e} + \mathcal{L}_{int}.
\eeq
Here $\lambda$ and $\alpha$ are coupling constants, which we will treat as free parameters.
The final term contains interactions, possibly strong, among the new particles so that none of the $Z$s are stable.  We will require $M \gg m > 2m_e$, and for simplicity we will take $m_i\gg M$ so that the $Z_i$s decay during the very early universe and their dynamics do not enter in the discussion below.  The only interaction with the standard model fields is the coupling between the $Y$ and electrons, which causes the $Y$s to decay at a rate $\Gamma_{Ye^+e^-}$. Note that  the coupling $\alpha$, which we take to be very small,  is not gauge invariant under the electroweak transformations but could easily arise in a low energy effective theory below the scale of electroweak symmetry breaking. We neglect  for now presentation of a UV complete  model, which would address, for instance, other possible couplings and decays of the $Y$ boson.

Under these conditions, the $X$ particle is stable and a WIMP dark matter candidate.  The key quantity in both the early and late universe is the annihilation rate of $X$s to $Y$s.  The total cross-section for this process is given by
\be
\sigma=\frac{\lambda^2}{8\pi s}\left( \frac{s-4m^2}{s-4M^2} \right)^\frac{1}{2}
\ee
where $\sqrt{s}$ is the center-of-mass energy.  At temperatures well below $M$, the energy of the incoming $X$s is
\be
\sqrt{s}\approx 2M+3T_X
\ee
where $T_X$ is the temperature of the $X$ particles.  To lowest order, the low-energy cross-section becomes
\be
\sigma \approx \frac{\lambda^2}{32\pi M^2}\left( \frac{M^2-m^2}{3MT_X} \right)^\frac{1}{2}.
\ee

\section{Early Universe}

The evolution of this model in the early universe is punctuated by several events.  Note that the temperatures of the hidden and standard sectors are generally unequal after they decouple, which takes place at some early time $t_{\mathrm{decouple}}$. We will assume that the $Z_i$ particles are heavy and have annihilated or decayed between times $t_{\mathrm{decouple}}$ and $t_{XY}$, the time when the annihilation of $X$ particles into $Y$ particles goes out of thermal equilibrium.  The main role of the $Z$ particles is to allow for the possibilility that  the temperature in the hidden sector $T_X$ is significantly higher than the temperature in the visible sector $T_\gamma$ at time  $t_{XY}$, which will turn out to enhance the relic abundance of $X$ particles today. The relic density of $X$ particle dark matter is determined by the decoupling of $X$ and $Y$ at time $t_{XY}$, which we assume to be later than $t_{\mathrm{decouple}}$.   We assume the    $Y$   particles are out of equilibrium with the standard model until time $t_{\mathrm{Ydecay}}$, given by the lifetime for $Y$ decays, at which time  the SM and the $Y$ particles come back into thermal equilibrium.   We will use the sudden decay approximation in which the reequilibration between the hidden and visible sectors occurs instantaneously. To avoid affecting the predictions of nucleosynthesis, we require that all these events occur before  the  nucleosynthesis epoch.

  After $X$ decouples from $Y$,  the $X$ momentum simply redshifts with expansion, while the number of $X$ particles per co-moving volume is approximately conserved. Similarly, the entropy per co-moving volume is nearly conserved, except   at time $t_{\mathrm{Ydecay}}$. It is thus convenient to compute the  ratio of $X$ particle density to entropy density. Once the increase in entropy density during the $Y$ decay epoch is computed, we can  determine  with the dark matter to photon ratio today.  

 We will begin with an approximate computation of  the observed present-day abundance of dark matter  in terms  of    the relic $X$ energy density  at $t_{XY}$. After $t_{XY}$, the dark matter energy density  evolves as $1/R^3$, where $R$ is the cosmological scale factor. Note that $X$ particles are  nonrelativistic at decoupling, and we take the mass of $X$ to be   constant.    Assuming    adiabatic evolution (no out of equilibrium processes)  between $t_{XY}$ and $t_{\mathrm{Ydecay}}$, we can use entropy conservation to relate  the photon temperature $T_{\gamma,XY}$ at $t_{XY}$ to the photon temperature $T_{\gamma,<\mathrm{Y decay}}$ just before $t_{\mathrm{Ydecay}}$
\be
T_{\gamma,<\mathrm{Ydecay}} =\frac{R_{XY}} {R_{\mathrm{Ydecay}}}
\left( \frac{ g_{*, XY} }{ g_{*,\mathrm{Ydecay}} } \right)^{\frac{1}{3}} T_{\gamma,XY}\ ,
\ee
 where $ {R_{XY}}/ {R_{\mathrm{Ydecay}}}$ is the ratio of the scale factor   at $t_{\mathrm{Ydecay}}$ to the scale factor at  $t_{XY}$, $g_{*, XY}$ is the effective number of relativistic species in equilibrium with the photon at time $t_{XY}$ and $g_{*,\mathrm{Ydecay}} $ is the effective number of relativistic species in equilibrium with the photon at time $t_{\mathrm{Ydecay}}$. At time $t_{\mathrm{Ydecay}}$, using the sudden decay approximation, the photons get reheated to a temperature $T_{\gamma,>\mathrm{Y decay}}$. We define a parameter $s$, the fractional change in the  entropy in the visible sector at $t_{\mathrm{Ydecay}}$  as
\be s\equiv \left(\frac{T_{\gamma,>\mathrm{Y decay}}}{T_{\gamma,<\mathrm{Y decay}}}\right)^3 \ .\ee 
 Note that $s$  in general satisfies the inequality
 \be\label{sineq} s>\frac{\left( T_{\gamma,<\mathrm{Y decay}}^4+\frac{1}{ g_{*,\mathrm{Y decay}} }T_{\mathrm{hid},\mathrm{Ydecay}}^4\right)^{\frac{3}{4}}}{T_{\gamma,<\mathrm{Ydecay}}^3}\ , \ee
where $T_{\mathrm{hid},\mathrm{Ydecay}}$ denotes the temperature the hidden sector.  The inequality is nearly saturated when the $Y$ bosons are relativistic when they decay. 
  
 We can now find the ratio  of $\rho_X (t_{XY})$,   the dark matter density at $t_{XY}$  to  $\rho_X (t_{0})$, the dark matter density today.
 \beq
  \frac{\rho_X (t_{0}) }{\rho_X (t_{XY})}&=&   \left( \frac{ g_{*,0} }{ s g_{*,XY} } \right) \left(\frac{T_{\gamma,0}}{T_{\gamma, XY}}\right)^3\cr &=&\left( \frac{ g_{*,0} }{ s g_{*,XY} } \right) \left( \frac{ T_{\mathrm{hid},XY}}{ T_{\gamma, XY} } \right)^3 \left(\frac{T_{\gamma,0}}{T_{\mathrm{hid},XY}}\right)^3\ .
\eeq
 This calculation differs from a standard      calculation of the WIMP abundance today by   the factors of $s$ and \be\left( \frac{ T_{\mathrm{hid},XY}}{ T_{\gamma, XY} } \right)^3 =\frac{g_{*,XY}}{g_{*,\mathrm{Ydecay}}}\left( \frac{ T_{\mathrm{hid},<\mathrm{Y decay}}}{ T_{\gamma,<\mathrm{Y decay}}} \right)^3.\ee Note that  we can use eq.\ref{sineq} to find an upper bound on the relic density compared with a standard relic abundance calculation in which all particles except the WIMP are in thermal equilbrium calculation. We define  $F$ such that 
 \beq F&\equiv& \frac{g_{*,XY}}{s g_{*, \mathrm{Ydecay}}}\left( \frac{ T_{\mathrm{hid},\mathrm{Y decay}}}{ T_{\gamma,<\mathrm{Y decay}}} \right)^3\cr
& \le&\frac{g_{*,XY}}{g_{*,\mathrm{Ydecay}}} \left( \frac{ T_{\mathrm{hid},\mathrm{Y decay}}}
{ \left( 
T_{\gamma,<\mathrm{Y decay}}^4+\frac{T_{\mathrm{hid},\mathrm{Y decay} }^4}{g_{*,\mathrm{Y decay}} }\right)^{\frac{1}{4}}
 } 
\right)^3\cr
& \le& \frac{g_{*,XY}}{g_{*,\mathrm{Ydecay}}^{\frac{1}{4}}}\ .\eeq
Saturating the inequality requires that the temperature in the hidden sector is much larger than the temperature of the visible sector, and that the $Y$ particles decay while relativistic.  If we take the photon temperature at the time of $XY$ decoupling to be between  5 and 80 GeV, then $g_{*,XY}$ only includes standard model particles lighter than the weak bosons, and we may take $g_{*,XY}=86\frac{1}{4}$. Assuming the $Y$ particles decay when the photon temperature is between 10  and 100 MeV  gives $g_{*,\mathrm{Ydecay}}=11\frac{3}{4}$ (including the contribution of the $Y$ particles). An approximate upper bound on $F$ is therefore
\be F < 46.6 \ . \ee
 The $X$ energy density at $t_{XY}$ is
\be
\rho_{X,XY}= M_X n_X \ee  where  
\be
n_X=\left( \frac{M_X T_{\mathrm{hid},XY}}{2\pi} \right)^{\frac{3}{2}}\exp(-M_X/T_{\mathrm{hid},XY}).
\ee 
This may be related to the dark matter energy density $\rho_{X,0}$ today as
\be \rho_{X,XY} =\frac{\rho_{X,0}}{ F }\frac{g_{*,XY}}{g_{*,0}}\left(\frac{T_{\mathrm{hid},XY}}{T_{\gamma,0}}\right)^3 \ .
\ee
The current temperature of the photons, $T_{\gamma,0}\approx 2.35\times 10^{-13}$ GeV, and the total present dark matter abundance is $\Omega_X\approx 0.11 h^{-2}$, where $h$ is the Hubble constant in units of $100 ({\rm km}/{\rm s})/{\rm Mpc}$.  Taking the favored value $h\approx 0.70$, the energy density of dark matter today is $\rho_{X,0} \approx 5\times 10^{-6} \mathrm{GeV} \mathrm{cm}^{-3}$.   For a given value of he model dependent factor $F$ we can relate $M$ and $T_{\mathrm{hid},XY}$. Note that the relation we  differs from a standard decoupling calculation by the factor of $F$.

\begin{figure}[htpb]
\includegraphics[width=8cm]{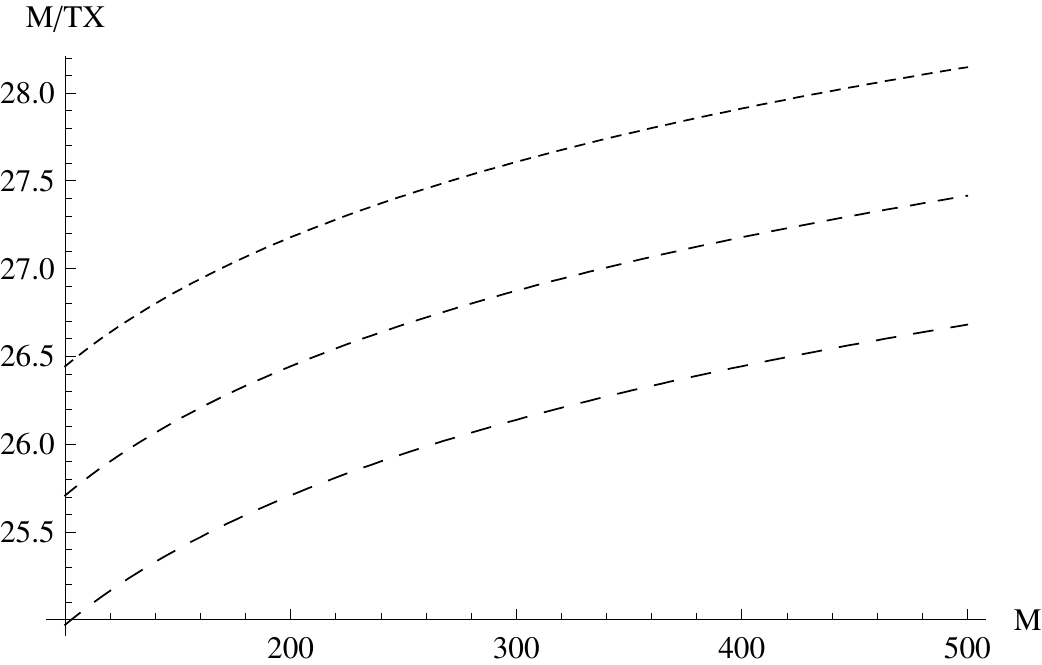}
\caption{The ratio of dark matter mass $M$ to its temperature at decoupling $T_{\mathrm{hid}, XY}$ for $F=46.6/i$ where $i=$\{1,2,4\} from top to bottom.}
\label{fig:decoupletemps}
\end{figure}

We plot the ratio of the mass and temperature $T_{\mathrm{hid},XY}$ of $X$ at freezout in figure \ref{fig:decoupletemps} for a number of  values of $F$, assuming the maximal value of $g_{*,XY}=86\frac{1}{4}$.


\section{Dark Matter in the Galaxy}

After structure formation occurs, dark matter is gravitationally pulled into a structures around galaxies called halos.  The density of the halo grows toward the center of the galaxy, and values are well above the mean value $\rho_X$.  The annihilation rate,
\be
\Gamma=n\sigma v, 
\ee
scales with the square of the number density $n$  and so significantly increases in these regions. For a non relativistic $v$, $\sigma v$ is approximately independent of $v$.  In this section we will calculate the resulting electron and positron spectrums and compare to observations made by PAMELA and ATIC.
Note that for a thermal relic,   $\sigma v$ is approximately determined by requiring that  
\be n_X\sigma v\approx \left(\frac{4 \pi^3 G_N(T_{\mathrm{hid},XY}^4+g_{*,XY}T_{\gamma,XY}^4)}{90}\right)^{\frac{1}{2}} .\ee
In a standard WIMP annihilation calculation where all other particles are in thermal equilibrium with the photons at decoupling the observed dark matter abundance today is obtained for
\be \sigma v\approx 3\times 10^{-26} \frac{\mathrm{cm}^3}{\mathrm{sec}}\ .\ee In our scenario this is modified by a factor of 
\be \sigma v\approx \frac{F(T_{\mathrm{hid},XY}^4+g_{*,XY}T_{\gamma,XY}^4)^{\frac{1}{2}}}{  {g_{*,XY}^{\frac{1}{2}}T_{\mathrm{hid},XY}^2}}3\times 10^{-26} \frac{\mathrm{cm}^3}{\mathrm{sec}}\ .\ee
Note that $F$ is largest when $T_{\gamma,XY}$ can be neglected relative to $T_{\mathrm{hid},XY}$.
The factor $F/ g_{*,XY}^{\frac{1}{2}}$ has an upper bound of about 5 so the upper bound on the annihilation cross section is approximately
\be  \sigma v<1.5 \times 10^{-25} \frac{\mathrm{cm}^3}{\mathrm{sec}}\ .\label{eq:svbound} \ee
A larger annihilation cross section would imply an upper bound on the temperature of the hidden sector which is below $T_{\mathrm{hid},XY}$ together with some nonthermal production process for $X$ particles, such as direct production in inflaton decay, or the mechanism in  ref. \cite{Fairbairn:2008fb}.

 For the sake of concreteness we will focus on the positron spectrum. 
The relevant process  for positron production is $XX\rightarrow YY \rightarrow 2e^+e^-$.  We fix the branching fraction for decays of the  intermediary $Y$ particles  to electron positron pairs to unity.  
 The flux at a given point in the galaxy are found by solving the diffusion equation of the positron density,
\be
\dot{f}-K\cdot \bigtriangledown^2f-\frac{\partial}{\partial E}(b f)=Q.
\ee
where $K$ and $b$ are energy-dependent diffusion and energy loss coefficients, and Q is the positron source term.  The procedure to obtain the electron spectrum is analogous.  A set of solutions to the diffusion equation in the case of a static distribution with disappearing flux outside a cylindrical region are well-known in the literature. A concise summary appears in \cite{Cirelli:2008id}.  For points near the solar system, the solution for the positron spectrum is
\be
\Phi(E)=B\frac{v_{e^+}}{4\pi b(E)}\frac{1}{2}\left(\frac{\rho_\odot}{M}\right)^2\int_E^MdE^\prime f(E^\prime)I(\lambda_D(E,E^\prime)).
\ee
In this expression $B$ is a dimensionless ``boost" factor that accounts for sub-halo structure, $v_{e^+}$ is the positron velocity, $b(E)$ is the energy loss coefficient, $\rho_\odot$ is the dark matter density near the solar system, and $I$ is the halo function defined in terms of the diffusion length $\lambda_D$.  $f(E)$ is the injection spectrum, given by,
\be
f(E)=\sigma v\frac{dN_{e^+}}{dE_{e^+}},
\ee
where $dN_{e^+}/dE_{e^+}$ for our model will be calculated below.

Recall that to leading order the cross-section for $XX\rightarrow YY$ has temperature dependence $\sigma \propto T_X^{-\frac{1}{2}}$.  The temperature cancels with the temperature dependence of the velocity, $v_X\propto T_X^{\frac{1}{2}}$, leaving the product $\sigma v$ temperature independent to leading order.

We now calculate the positron injection spectrum $dN/dE$ created by $X$ annihilation.  In the galactic frame, the total energy of the colliding $X$ particles is $\sqrt{s}=2M$, so the ejected relativistic $Y$ particles have the Lorentz factor
\be
\gamma_Y=(1-\beta_Y^2)^{-1/2}=\frac{M}{m}.
\ee

\begin{figure}[htpb]
\includegraphics[width=8cm]{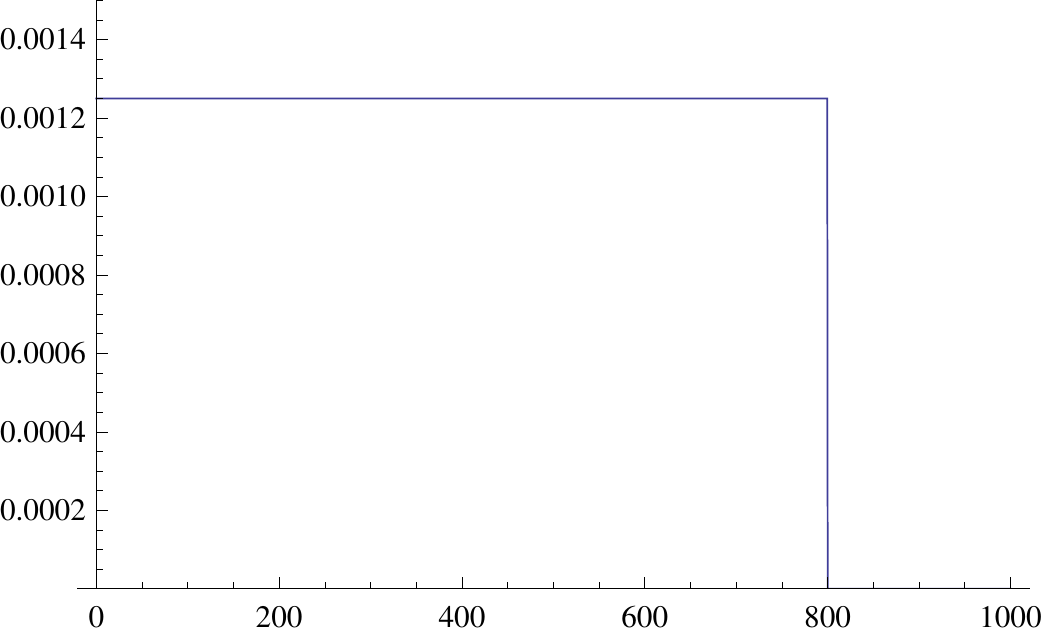}
\caption{The energy spectrum dN/dE of the electrons as a function of the lab frame energy for M=800GeV, m=200MeV}
\label{fig:injection}
\end{figure}

In the center of mass frame of Y, it decays to $e^+e^-$ isotropically and monochromatically.  Boosting the positrons back to the galactic frame yields an energy of
\be
E_{e^+}=\frac{M}{2}-\frac{\sqrt{M^2-m^2}}{m}\sqrt{\frac{m^2}{4}-m_e^2}\cos\theta
\eqn{eq1}
\ee
where $\theta$ is the angle between the $Y$ spatial momentum and positron spatial momentum.
Since the emission of positrons in the frame of $Y$ is isotropic, the above relation is readily converted to the injection spectrum by
\be
\frac{dN_{e^+}}{dE_{e^+}}=\frac{d\cos\theta}{dE_{e^+}}\frac{dN_{e^+}}{d\cos\theta},
\ee
where $dN_{e^+}/d\cos\theta=1/2$.  The result is
\beq
\frac{dN_{e^+}}{dE_{e^+}} = \frac{m}{2\sqrt{M^2-m^2}\sqrt{m^2-4m_e^2}}
\eeq
which is shown in figure \ref{fig:injection}.

There are two sets  of parameters needed for a calculation of the observed positron and electron spectrums.  The first are the parameters of the model we have described above. Assuming the dark matter consists of $X$ particles, the relevant parameters are   $M$, $m$,  and $\lambda$.  The second set describe the mass distribution of the dark matter halo, and the diffusion of positrons and electrons.  The parameters of the second set which are not known with precision are $B$ and the factors that enter the diffusion length, $K_0$ and $\delta$.  For the latter set, we will study the exemplar models from recent dark matter literature, which are outlined in \cite{Cirelli:2008id} and \cite{Delahaye:2007fr}.  We considered three standard halo profiles, NFW, Moore and IsoT, and three positron diffusion models, denoted ``min," ``med," and ``max".

In practice, there is little dependence on the choice amongst the profile models, and here we will only present results from the NFW profile model.  In addition the ``med" and ``max" propagation models yield very similar spectrums at Earth.  The cross-section, and hence the flux, has very little dependence on the value of the $Y$ mass $m$ provided that it satisfied the kinematic bound on electron production.  We will fix its value at 200 MeV.

The values of $B$ and $\lambda$ are degenerate in generating dark matter annihilation fluxes.  The flux $\Phi$ depends only on an overall scaling of $B\lambda^2$.  We will require that the boost factor is within the bounds usually discussed in the literature, less than 20.  Some studies suggest that even values of $B$ above 2-3 are disfavored~\cite{Lavalle:1900wn}.  We will also require that $\lambda$ is sufficiently small that the $Y$-electron coupling remains perturbative, which is true up to values of $\lambda \lsim 10$.  We will find that we can generate spectra well above background levels without approaching either of these bounds.  

\begin{figure}[htpb]
\includegraphics[width=8cm]{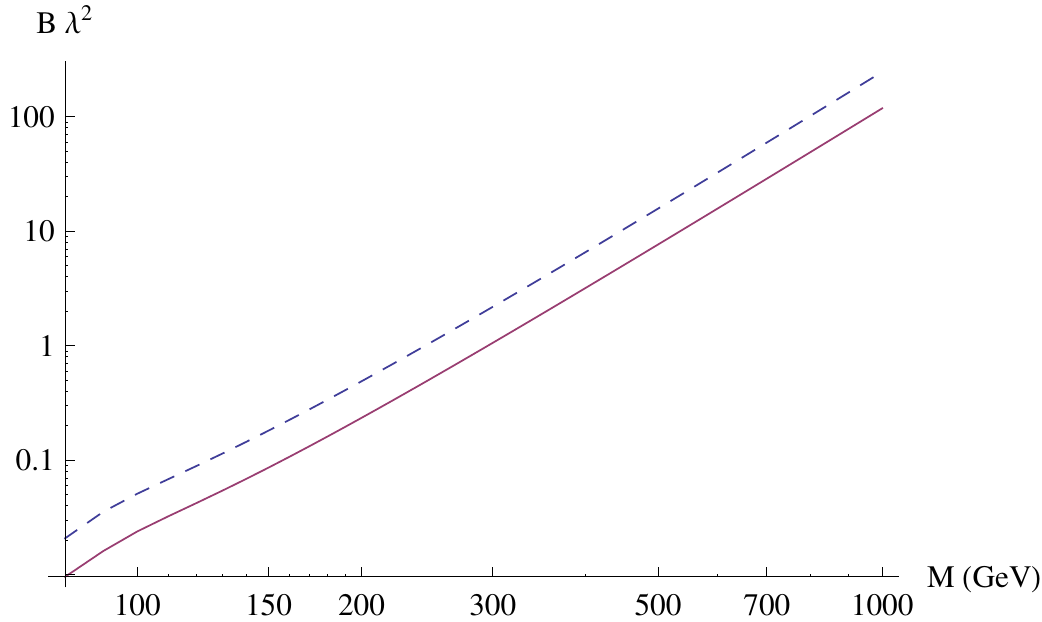}
\caption{Required values of $B\lambda^2$ to generate the positron excess described in the text.  The top dashed curve is the ``min" model, and the solid curve is the ``med" models.}
\label{fig:bvals}
\end{figure}

We fit our model to the recent PAMELA and ATIC observations, both individually and simultaneously.  The PAMELA data which has been publicly discussed so far was presented as the ratio of the total number of positrons to the total number of positrons and electrons.  This ratio departs background levels at 10 GeV, and rises to a value of 10\% at 60 GeV.  We fit the required value of $B\lambda^2$ required to generate such a spectrum as a function of $M$.  The result is presented in figure~\ref{fig:bvals}.  

Here there is some slight tension between the parameter $B$, $\lambda$ and the cross-section required to generate the relic abundance.  Smaller $B$ values require larger $\lambda$, however this pushes us against the bound on $\sigma v$ required to generate the present abundance from a thermal relic given in eq.~(\ref{eq:svbound}).  For an 800 GeV $X$ particle and the ``med" propagation model, the fit produces $B=4.6$ for $\lambda=4$.  However, these values yield $\sigma v=1.4\times 10^{-24} \cm^3/{\rm s}$.  We hit our bound of $B=20$ for $\lambda=1.9$, which produces $\sigma v=3.3\times  10^{-25} \cm^3/{\rm s}$.  This tension is significantly relaxed for smaller masses $M$.

\begin{figure}[htpb]
\includegraphics[width=8cm]{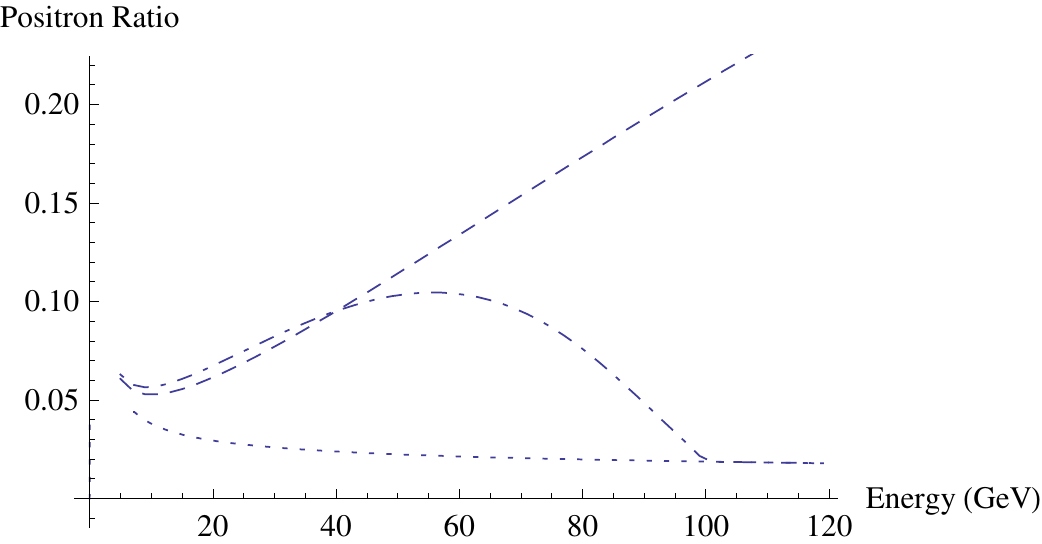}
\caption{Positron excess below 120GeV.  The dash-dot curve is $M=100\GeV$.  The dashed curve is $M=800\GeV$.  The bottom dotted line is the background level.}
\label{fig:pamfits}
\end{figure}

Spectra for two values of $M$ are shown for the lower half of the PAMELA energy range in figure~\ref{fig:pamfits}, where the $B$ factor has chosen to be optimal.  The curves for two values of $M$ are nearly degenerate below 60 GeV.  Lower values of $M$ show distinct spectral shapes that would be easily distinguished in the full PAMELA data release.  Higher values of $M$ have nearly identical shapes in the PAMELA range.

We now turn our attention to the ATIC results.  The collaboration presented a spectrum in terms of total electron flux that exhibits an excess well above their expected background level.  The excess begins at 300 GeV and peaks at about 500 GeV.  Their data at 900 GeV and above show no evidence for an excess.  At the peak, the total number of electrons is around twice the background level.

\begin{figure}[htpb]
\includegraphics[width=8cm]{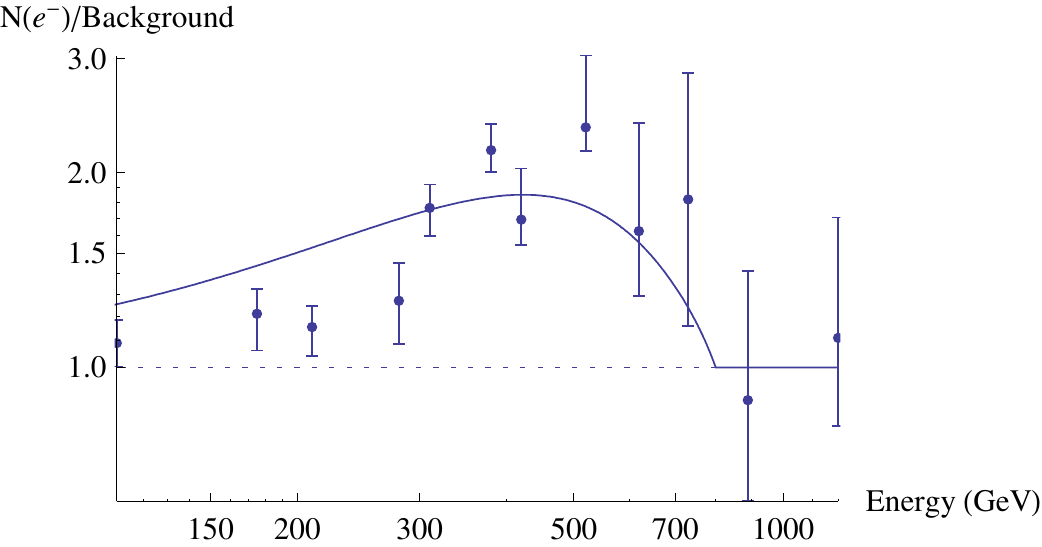}
\caption{The ratio of the ATIC signal to background along with the model fit for the parameters described in the text.}
\label{fig:aticfit}
\end{figure}

We fit the ATIC data in terms of the ratio of the total number of electrons to the expected background number.  We find that the excess is well fit for values of $M\gsim 700\GeV$.  Remarkably, the best-fit boost values are quite similar to those required by the PAMELA data for masses in this range.  The ATIC spectrum, using $M=800\GeV$ and the best-fit PAMELA boost factor, is shown in figure~\ref{fig:aticfit}.

\begin{figure}[htpb]
\includegraphics[width=8cm]{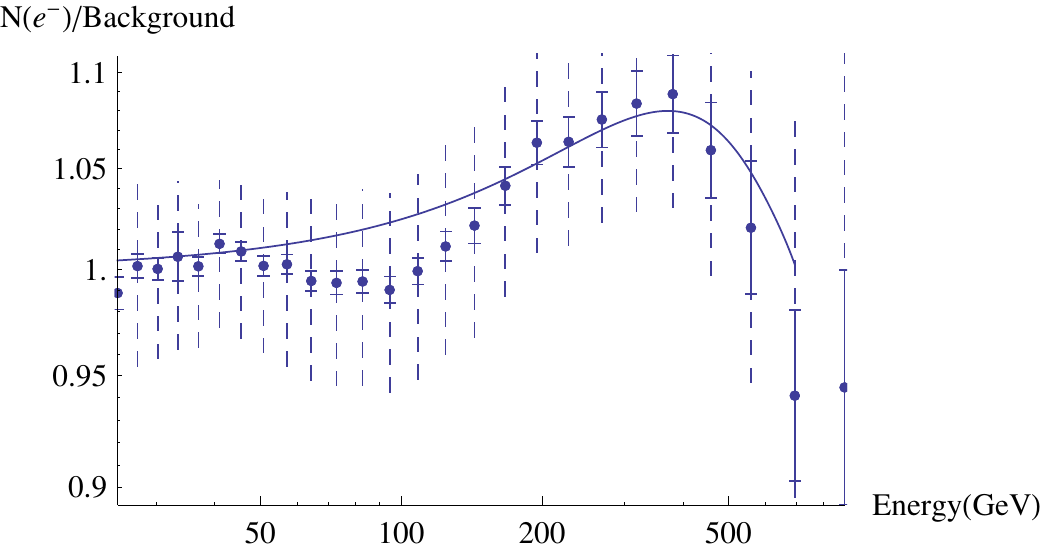}
\caption{The ratio of the Fermi signal to background along with the spectrum for the dark matter model with parameters described in the text. Solid error bars indicate statistical errors, and dotted lines indicate the sum of statistical and systematic errors.}
\label{fig:fstfit}
\end{figure}

We also fit the Fermi electron data \cite{Abdo:2009zk} to our model. To estimate the background, we make a power law fit to the Fermi points below 100 GeV, below the bump feature. We then rescale the spectrum to this background, as we had for ATIC, and find the best-fit balues for $M$ and boost. We found reasonable results for $M=700\GeV$ and boost $B\lambda^2=0.4$. The resulting spectrum is shown in figure~\ref{fig:fstfit}. The spectrum generated by our model is softer than the Fermi spectrum when only statistical errors are considered, however the agreement greatly improves when systematic errors are included. Note that there is tension between the parameters which give a best fit to Fermi data and those that give the best fit to PAMELA / ATIC data in figure \ref{fig:bvals}. Such tension is common to all models of direct annihilation of sub-TeV dark matter into messenger particles which decay into $e^+e^-$ pairs, and can be reduced or eliminated by assuming a larger uncertainty in the Fermi energy resolution \cite{Hooper:2009cs}.

\section{Conclusion}

Despite the compelling case for dark matter, we have few clues as to its nature, so it is important to consider a variety of possibilities with distinctive signatures. A nonminimal dark sector, consisting of more than one new particle interacting among themselves relatively strongly,    occurs in a number of interesting scenarios, such as hidden valley models \cite{Strassler:2006im} and unparticle models \cite{Georgi:2007ek}. In such scenarios, a stable WIMP may annihilate into lighter unstable particles in  the dark sector. The   cross section for such annihilation  can be   larger than in typical   WIMP models. The   cross section for direct detection of  dark matter would be very small, while decay of the lighter unstable particles into standard model particles offers new possibilities for indirect detection. Such enhanced indirect detection signals  are particularly interesting in light of the recent PAMELA and ATIC reports, as well as earlier reported excesses in the cosmic ray spectrum. In this paper we have considered the cosmology of a  simple example, where the light intermediary is a scalar particle, which may produce an  excess  of high energy positrons and electrons in cosmic rays.   We have shown that the model may easily explain the PAMELA and ATIC data. If the scalar coupling to fermions is proportional to fermion mass, as would be expected if this coupling arises from mixing with the Higgs, electron decay  dominance would imply that the scalar is lighter than about 200 MeV. Decays to 2 photons mediated by loops of heavier fermions would also be expected, giving a     gamma ray signal, possibly visible in  EGRET  and FERMI gamma ray observatory data \cite{Sreekumar:1997un,Atwood:1993zn}. Such  a gamma ray spectrum would distinguish our model from those with a light vector intermediary \cite{Pospelov:2007mp,Pospelov:2008jd,ArkaniHamed:2008qn}, as a  vector particle cannot decay to 2 photons, and decays to 3 photons would be typically expected to have a small branching fraction.  In addition, a scalar would not decay to neutrinos, so indirect detection would not be possible in neutrino observatories \cite{Achterberg:2006md}. 
In  the minimal version of the model we have presented, the dark sector would not show up in collider experiments,  although it would be simple to extend  the model so that  additional new particles with stronger standard model interactions could be produced at the LHC whose decay  products include $X$   or $Y$ particles. 
\section*{Acknowledgments}
This work was supported in part by the DOE under contract DE-FG02-96ER40956.
We thank Torsten Bringmann, Marco Cirelli, and  Neal Weiner for helpful comments on an earlier version of the manuscript and for bringing important references to our attention.
\bibliographystyle{apsrev}
\bibliography{dm-hl}

\end{document}